\documentclass[11pt,twoside]{article}
\usepackage{./asp2010}

\resetcounters

\begin{document}

\title{Very Massive Stars and the Eddington Limit}
\author{Paul A. Crowther$^1$, R. Hirschi$^{2}$, Nolan R. Walborn$^{3}$,
and Norhalisza Yusof$^{2,4}$
\affil{$^1$Dept of Physics \& Astronomy, University of Sheffield, 
Hounsfield Rd, Sheffield, S3 7RH, UK}
\affil{$^2$Astrophysics Group, EPSAM, University of Keele, Lennard-Jones 
Labs, Keele, ST5 5BG, UK}
\affil{$^3$Space Telescope Science Institute, 3700 San Martin Drive, 
Baltimore, MD 21218, USA}
\affil{$^{4}$Department of Physics, University of Malaya, 50603 
Kuala Kumpur, Malaysia}
}

\begin{abstract}
We use contemporary evolutionary models for Very Massive Stars (VMS) 
to  assess whether the Eddington limit constrains the upper stellar
mass  limit. We also consider the interplay between mass and age
for the wind properties and spectral morphology of VMS, with
reference to the recently modified classification scheme for 
O2--3.5\,If*/WN  stars. Finally, the death of VMS in the local universe 
is considered in the context of pair instability supernovae.
\end{abstract}

\section{Eddington limit}

Empirical determinations for the upper stellar mass limit, 
$M_{\rm max}$, 
have led to the adoption of $M_{\rm max} \sim 150 M_{\odot}$
\citep[e.g.][]{Figer05, OeyClarke05}. 
This limit closely coincides with the `first approximation' for
$M_{\rm max}^{\rm Edd} \sim 200 M_{\odot}$, i.e. the
intersection between the Eddington limit and 
the $L \propto M^{3}$ mass-luminosity relation 
for main sequence stars
\citep[e.g.][his Fig 3.6]{Maeder09}. Therefore, two obstacles need 
to be overcome for these findings to be reconciled with the claimed
320 $M_{\odot}$ initial mass for R136a1 by \citet{Crowther10}.
\citet{Massey11} has provided convincing arguments 
regarding the empirical estimates of $M_{\rm max}$, 
so here we focus attention on $M_{\rm max}^{\rm Edd}$.

\begin{figure}
\begin{center}
\includegraphics[width=7cm]{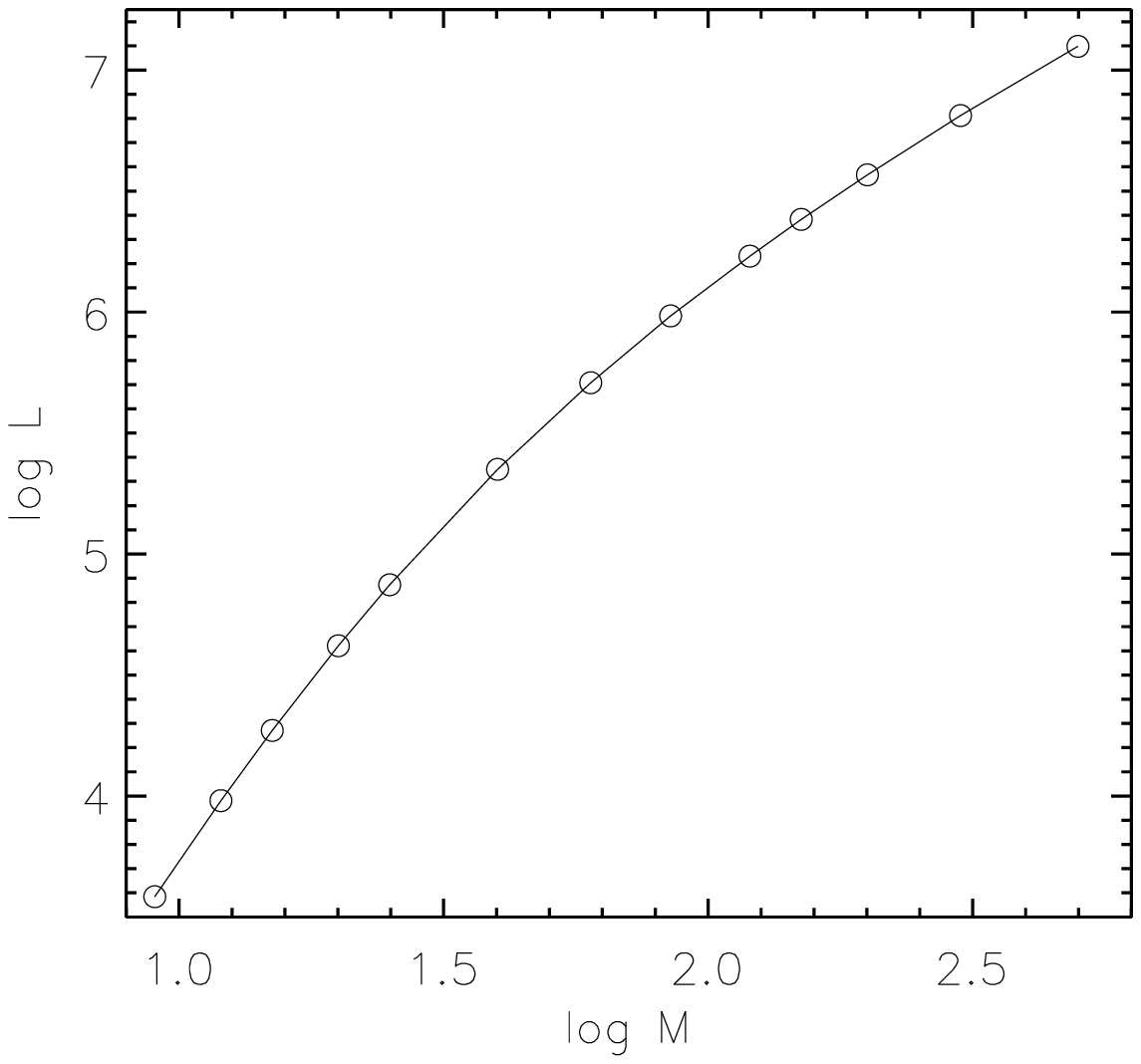}
\caption{Mass-luminosity relation for non-rotating, solar metallicity 
ZAMS stars from \citet[][$\leq 85 M_{\odot}$]{MeynetMaeder00} 
and \citet[][$\geq 120 M_{\odot}$]{Crowther10}.}
\label{fig1}
\end{center}
\end{figure}

Main sequence models for very massive stars (VMS) have been calculated 
using 
the Geneva evolutionary code by R. Hirschi and N. Yusof \citep[see][]{Crowther10} and the 
Bonn evolutionary code by K. Friedrich \citep[see][]{Graefener11}.  
In Figure~\ref{fig1} we present the mass--luminosity 
relation for non-rotating, solar metallicity ZAMS, 
incorporating 9-85 $M_{\odot}$ models from \citet{MeynetMaeder00}. 
Although $x$ = 3  represents a sensible average for $L \propto 
M^{x}$  across all stellar masses, $x \sim$ 2.5 for 10--20 
$M_{\odot}$ and flattens further at higher masses, reaching $x \sim$ 
1.5 close to 200 $M_{\odot}$. 


The Eddington parameter, $\Gamma_{e}$,  can be expressed as 
\[ \Gamma_{e} = g_{e}/g = 3\times10^{-5} q \frac{L/L_{\odot}}{M/M_{\odot}} \]
where $q$ = 0.86 for main sequence hot stars. A decrease in the slope of 
the mass-luminosity relation at very high masses reduces $\Gamma_{e}$,
and so raises $M_{\rm max}^{\rm Edd}$. Of course, $\Gamma_{e}$
increases once a star evolves away from the ZAMS ($L/M$ increases),
and since $\dot{M} \propto L (\Gamma_{e}/(1-\Gamma_{e}))^{2/3}$
for radiatively driven winds with a CAK power index of $\alpha \sim$ 0.6 
\citep{Owocki03}, stronger winds  
are anticipated  both qualitatively \citep{SmithConti08} and 
quantitatively \citep{GraefenerHamann08} with age.

\begin{figure}
\begin{center}
\includegraphics[width=7cm]{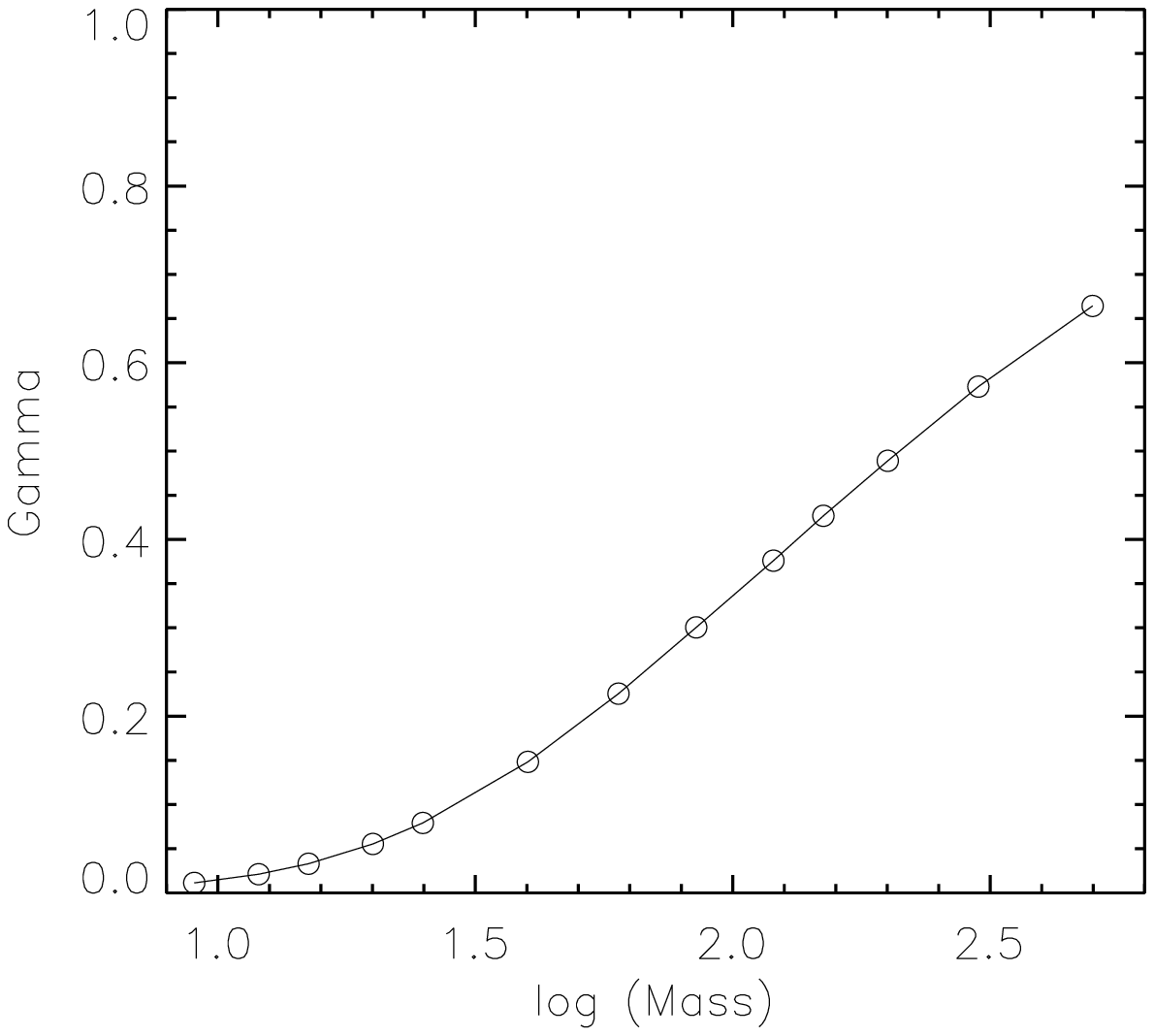}
\caption{Eddington parameter, $\Gamma_{e}$, for non-rotating, solar 
metallicity ZAMS stars from \citet[][$\leq 85 M_{\odot}$]{MeynetMaeder00} 
and \citet[][$\geq 120 M_{\odot}$]{Crowther10}.}
\label{fig2}
\end{center}
\end{figure}

Figure~\ref{fig2} compares 
$\Gamma_{e}$ for ZAMS spanning 10--500 $M_{\odot}$ at solar composition. 
This illustrates 
that the Eddington limit is not approached, 
so $M_{\rm max}^{\rm Edd}\gg 500 M_{\odot}$. In fact, $x\rightarrow 1$ as $M \rightarrow \infty$, so
the Eddington limit might never be reached for ZAMS stars. Once on the main sequence, $\Gamma_{e}$, and in turn 
mass-loss rates, increase with both age and mass, so strong wind 
signatures may correspond either to a relatively evolved high mass 
star or an unevolved very high mass star. Table~\ref{table1} compares the 
influence of mass and age upon spectral type
for the case of the coeval cluster R136a whose age is $\sim$1.5 Myr 
\citep{Crowther10}. 

\begin{table}[!ht]
\caption{Influence of mass (vertical) and age (horizontal) upon spectral type
for the young LMC star cluster R136a}
\label{table1}
\begin{center}
\begin{tabular}{llll}
\tableline
\noalign{\smallskip}
Initial & Sp Type & Sp Type  & Example \\
Mass ($M_{\odot}$)   & (ZAMS)  & (1.5 Myr)&         \\
\noalign{\smallskip}
\tableline
\noalign{\smallskip}
240 & O2\,If*? & WN5h & R136a2 \\
140 & O2\,III?  & O2\,If* & R136a5 \\
100 & O2--3\,V? & O3\,III(f*) & R136a7 \\
50  & O3\,Vz?   & O3V         & [HSH95] 50$^{\ddag}$ \\
\noalign{\smallskip}
\tableline
\end{tabular}
\end{center}
$\ddag$: \citet[][HSH95]{Hunter1995}
\end{table}


\section{Transition Of/WN stars}

From the previous section, high $\Gamma_{e}$'s 
develop either in young, very high mass stars or evolved lower mass
stars. Spectroscopic signatures of strong winds in early type stars
include He\,{\sc 
ii} $\lambda$4686 and/or H$\alpha$ emission, corresponding to OBA supergiants
or Wolf-Rayet stars in the case of very strong emission features. 
From Table~\ref{table1}, the current spectral type of R136a2 is WN5h, while
its ZAMS spectral type may have resembled an O2 supergiant. Had we
witnessed R136 perhaps 0.5 million years ago, it would have exhibited
an intermediate spectral type. Indeed, a hybrid O3\,If*/WN category -- 
spectroscopically intermediate  between 
early  O stars and WN stars -- was introduced by \citet{Walborn82}. 

\begin{figure}
\begin{center}
\includegraphics[angle=-90,width=12cm]{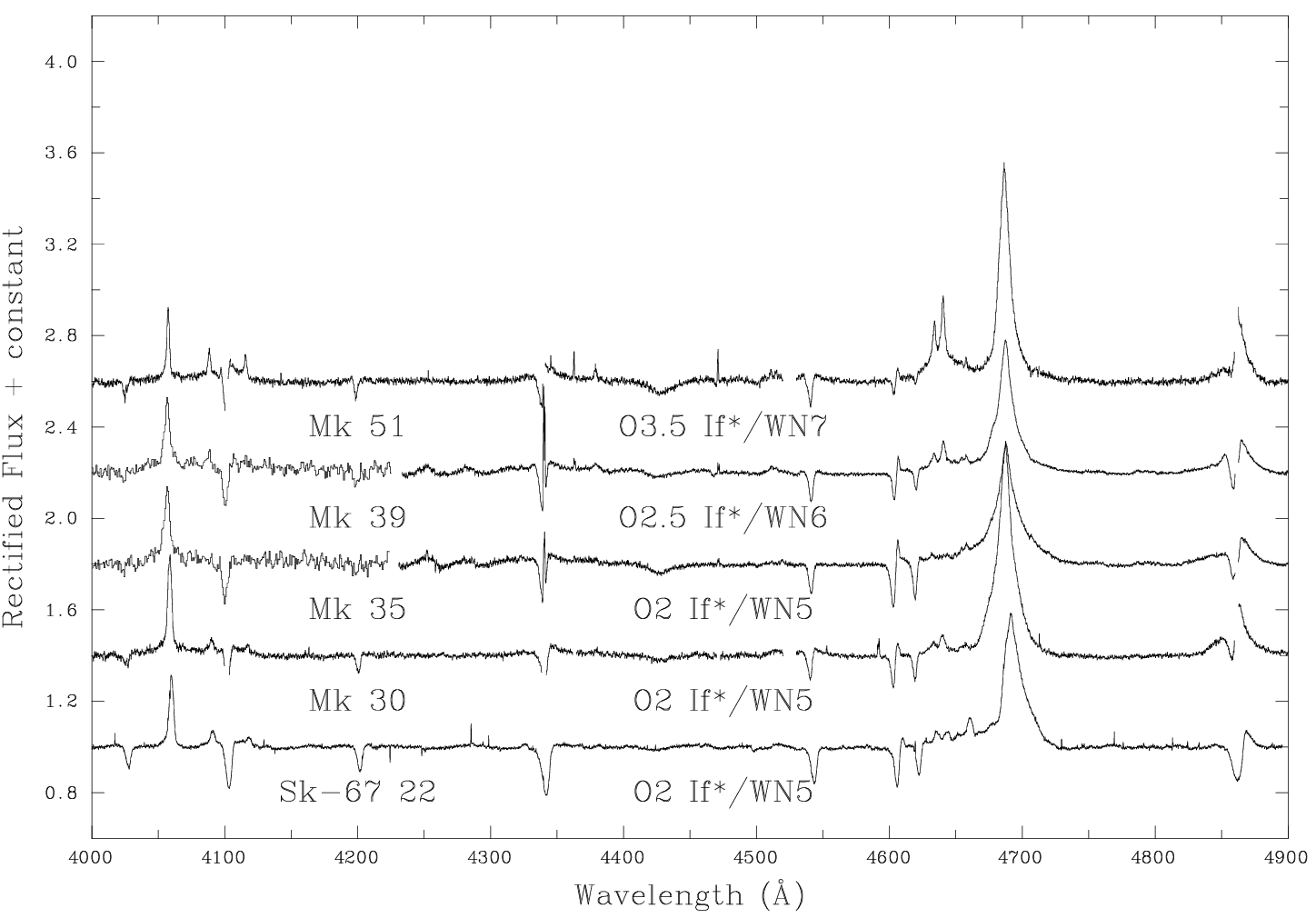}
\caption{Spectrograms of transition Of/WN stars \citep{CrowtherWalborn11}
}
\label{fig3}
\end{center}
\end{figure}

Following the extension of the MK sequence to O2 
\citep{Walborn02} and revisions to WN classifications, this has been 
refined recently by \citet{CrowtherWalborn11}.
Spectroscopically the morphology of H$\beta$ is key to O2--3.5\,If*,
O2--3.5\,If*/WN or WN subtypes, while a qualitative interpretation 
led \citet{CrowtherWalborn11} to conclude that most O2--3.5\,If*/WN stars 
(e.g. Melnick 35) are 
very  luminous, young stars with $150\pm 30 M_{\odot}$. However, some 
Of/WN  stars are substantially lower in luminosity/mass (e.g. Sk 
--67$^{\circ}$ 22, Melnick 51), with correspondingly larger ages, even
though these may be spectroscopically indistinguishable from other
examples, as illustrated in Fig.~\ref{fig3}.

The incidence of O2--3.5\,If*/WN stars in the LMC is significantly higher than
in the Milky Way. Radiatively driven winds of Galactic stars would be 
expected to be modestly higher than LMC counterparts, so one would 
predict a slightly higher percentage of O2--3.5\,If*/WN stars in the LMC with 
respect to the Milky Way. In fact, O2--3.5\,If*/WN 
stars comprise 7\% of the 106 
WN-type stars in the LMC \citep{Breysacher99}, versus only 2\% of 
the highly incomplete 175 WN stars compiled by \citet{vanderHucht01, vanderHucht06}. 
Since transition spectral types arise preferentially in very massive stars,
one would expect them predominantly in regions of the highest star 
formation. Indeed, the 30 Dor region of the LMC dominates Of/WN statistics 
in the Local Group \citep{CrowtherWalborn11}.

\section{Death of Very Massive Stars}

A natural question relating to VMS is whether they would follow the 
usual path to core-collapse supernovae (CCSNe) or explode prematurely as 
pair-creation super\-novae (PCSNe)? \citet{Heger03} concluded that
metal-free, single massive stars with 140--260 $M_{\odot}$ would 
explode as PCSNe. However, unbiased transient surveys have recently 
identified
exceptionally bright supernovae in the local universe, some of which have 
been attributed to PCSNe from initially $\sim 200 M_{\odot}$ stars 
with more modest metal-deficiencies \citep[e.g. SN 2007bi][]{GalYam09}.

\begin{figure}
\begin{center}
\includegraphics[width=9cm]{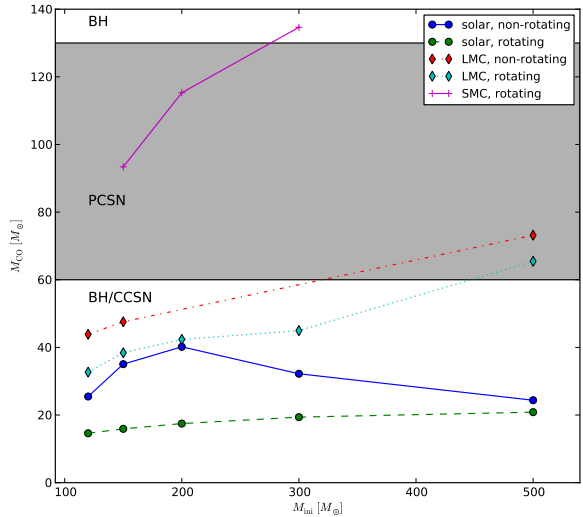}
\caption{Final CO mass versus initial mass for VMS at SMC, LMC
and solar metallicities, with the domain of PCSNe shown in grey,
based upon mass-loss prescriptions of \citet{Vink01} and \citet{NugisLamers00}
for, respectively, the H-rich and H-poor phases (from Yusof et al. in 
prep.).}
\label{fig4}
\end{center}
\end{figure}

VMS models have been calculated throughout their post-main sequence 
evolution using \citet{Vink01} mass-loss prescriptions for the main
sequence and \citet{NugisLamers00} for the post-main sequence Wolf-Rayet 
phase, the results of which are presented in Fig.~\ref{fig4}. 
H-deficient CCSNe are predicted for 100--300 $M_{\odot}$ stars at 
solar and LMC metallicities, whereas CO core masses of 60--130 
$M_{\odot}$  are obtained for rotating 150--200 $M_{\odot}$ stars at SMC 
metallicity. Therefore, VMS at low metallicity  may indeed produce PCSNe.
Indeed, \citet{Quimby11} have 
identified a class of luminous H-deficient supernovae located in 
faint, metal-poor host galaxies. 
\citeauthor{Quimby11} attributed such bright SNe
to a strong interaction between the CCSN of a very massive  star and a 
H-free shell produced by violent pulsations, that was perhaps initiated 
by the pair instability.

However, these predictions are very sensitive to mass-loss 
prescriptions, especially for the post-main sequence phase. For the
Wolf-Rayet phase separate expressions are adopted for WN and WC stars, 
in which mass-loss rates are expressed in terms of luminosity and
composition \citep[eqns. 20--21]{NugisLamers00}. Such calibrations, based on
results for Wolf-Rayet stars at solar composition, imply a factor of $\sim$30 
increase in mass-loss rate from the H-rich \citep{Vink01} to 
the H-deficient phase of a 300 $M_{\odot}$ star at SMC metallicity. 
To illustrate the sensitivity, let us alternatively adopt equation 25 from 
\citet{NugisLamers00}, albeit modified to allow for the  $\dot{M} \propto Z_{\rm Fe}^{0.7}$ dependence of mass-loss upon ambient (Fe-peak) 
metallicity, $Z_{\rm Fe}$
\citep{Crowther02, VinkDeKoter05, Crowther06}, i.e.
\[ \log \dot{M/(M_{\odot} {\rm yr}^{-1})} = -5.7 + 0.88 \log (M/M_{\odot}) + 0.7 \log (Z_{\rm Fe}/Z_{Fe, \odot}) \] 
This would exceed the \citet{Vink01} prediction by only a factor of 2
for the case of a SMC metallicity 300 $M_{\odot}$ star, leading to significantly
higher CO masses than those presented in Fig.~\ref{fig4}, raising 
the possibility of PCSNe from VMS progenitors at higher metallicity.

\acknowledgements Partial financial support was provided by the 
University of Sheffield's Alumni fund. Thanks to Chris Evans and 
the VLT FLAMES Tarantula Survey consortium for selected Of/WN spectroscopy, 
plus Norhasliza Yusof for individual VMS calculations.

\bibliography{crowther}

\question{{\bf Krysztof Stanek} Is there any photometric variability 
information for your most massive stars?} 

\answer{{\bf Paul Crowther} Not that we know, although their location at 
the centres of very crowded clusters makes photometric studies 
challenging (R136a1 and R136a2 are separated by only 0.1 arcsec)}

\end{document}